Giant tunnel magnetoresistance and high annealing stability in CoFeB/MgO/CoFeB magnetic tunnel junctions with synthetic pinned layer

Young Min Lee

*Laboratory for Nanoelectronics and Spintronics, Research Institute of Electrical Communication, Tohoku University, 2-1-1 Katahira, Aoba-ku, Sendai 980-8577, Japan*

Jun Hayakawa

*Hitachi, Ltd., Advanced Research Laboratory, 1-280, Higashi-koigakubo, Kokubunji-shi, Tokyo 185-8601, Japan, Laboratory for Nanoelectronics and Spintronics, Research Institute of Electrical Communication, Tohoku University, 2-1-1 Katahira, Aoba-ku, Sendai 980-8577, Japan*

Shoji Ikeda, Fumihiro Matsukura and Hideo Ohno

*Laboratory for Nanoelectronics and Spintronics, Research Institute of Electrical Communication, Tohoku University, 2-1-1 Katahira, Aoba-ku, Sendai 980-8577, Japan*

We investigated the relationship between tunnel magnetoresistance (TMR) ratio and the crystallization of CoFeB layers through annealing in magnetic tunnel junctions (MTJs) with MgO barriers that had CoFe/Ru/CoFeB synthetic ferrimagnet pinned layers with varying Ru spacer thickness ($t_{Ru}$). The TMR ratio increased with increasing annealing temperature ($T_a$) and $t_{Ru}$, reaching 361% at $T_a$ = 425°C, whereas the TMR ratio of the MTJs with pinned layers without Ru spacers decreased at $T_a$ over 325°C. Ruthenium spacers play an important role in forming an (001)-oriented bcc CoFeB pinned layer, resulting in a high TMR ratio through annealing at high temperatures.

There is a great deal of interest in using magnetic tunnel junctions (MTJs)[1-3] to realize non-volatile magnetoresistive random access memories (MRAM). For MRAM applications, MTJs must endure thermal treatment at the high temperature of 400°C, because such treatment is necessary in order to complete the Si integrated circuit processing, into which MTJs are integrated.[4] However, thermal instability in conventional MTJs with aluminum oxide tunnel barriers has been one of the problems in MRAM fabrication; the tunnel magnetoresistance (TMR) ratio rapidly decreases when the MTJs are annealed at temperatures over 280°C.[4-6] Recent advances in MTJs using (001)-oriented MgO barriers and CoFeB electrodes made it possible to achieve high TMR ratios through crystallization of initially amorphous CoFeB electrodes at annealing temperatures ($T_a$) as high as 375°C,[7-11] indicating that the MTJs using MgO barriers and CoFeB electrodes are more robust against high temperature annealing. The mechanism of high thermal stability, however, has not yet been fully understood in CoFeB/MgO/CoFeB MTJs.

Here, we investigated the relationship between the TMR ratio and the crystallization of the CoFeB layer through annealing in MTJs with highly (001)-oriented MgO barriers, CoFeB free layers, and CoFe/Ru/CoFeB synthetic ferrimagnet (SF) pinned layers. We varied the Ru spacer thickness ($t_{Ru}$) from 0.67 nm to 2.8 nm and compared them with MTJs without the spacer. The TMR ratio increased with increasing $T_a$ and $t_{Ru}$, reaching a maximum TMR ratio of 361% at the $T_a$ of 425°C. Crystallization to a highly oriented bcc (001) structure in the CoFeB pinned layer deposited on the Ru spacer was found to be the key in achieving the high TMR ratio through annealing at high $T_a$.

The MTJs used in this study are composed of Ta(5)/Ru(50)/Ta(5)/NiFe(5)/IrMn(8)/pinned layer/MgO(1.7)/CoFeB(3)/Ta(5)/Ru(15) (parenthetical units are in nm) deposited on Si/SiO$_2$ wafers with 3-inch diameters using RF magnetron sputtering. The MgO barrier was sputtered directly from a MgO target at 10 mTorr in an Ar atmosphere at the deposition rate of 0.015 nm/s. Other films were deposited at 1 mTorr in an Ar atmosphere. We use CoFe to represent a Co$_{50}$Fe$_{50}$ alloy and CoFeB to represent a Co$_{40}$Fe$_{40}$B$_{20}$ alloy throughout this work. (Here, nominal target compositions are used). Two types of pinned layers were prepared: an S(single)-MTJ type, composed of CoFe(2.5)/CoFeB(3) with no Ru spacer, and an SF(synthetic ferrimagnet)-MTJ type, composed of CoFe(2.5)/Ru(0.67–2.8)/CoFeB(3). The Ru spacer in the SF multilayer was fabricated using the slide shadow mask technique so that its thickness increased from 0.67 to 2.8 nm. A schematic illustration of the MTJ stacks is shown in Fig. 1. All prepared samples were micro-fabricated by photolithography with a junction size of 0.8×0.8 - 0.8×5.6 μm$^2$. The completed samples were annealed in a vacuum of 10$^{-5}$ Pa for 1 hour under an applied magnetic field of 4 kOe, varying the $T_a$ from 270°C to 450°C. We measured electrical properties of the samples using a dc four-probe method in the magnetic field range of ±3 kOe. High-resolution cross-sectional transmission electron microscopy (HRTEM) and x-ray diffraction (XRD) measurements were used to investigate the crystalline structure of samples. All measurements were carried out at room temperature. We define the TMR ratio as $(R_{ap}-R_p)/R_p \times 100$, where $R_{ap}$ and $R_p$ are the resistance at parallel (P) and antiparallel (AP) configurations of two magnetic electrodes, respectively. The exchange anisotropy field ($H_{ex}$) caused by exchange coupling between the IrMn and CoFe layers was defined by the external field at which the TMR ratio becomes half of its maximum value.

Figure 2 plots $H_{ex}$ (a) and TMR ratios (b) of the MTJ samples as a function of Ru spacer thickness for $T_a$ of 270°C, 375°C, and 425°C; the values at $t_{Ru} = 0$ represent those for S-MTJ. As seen in Fig. 2 (a), oscillation in $H_{ex}$ with respect to the Ru spacer thickness is observed at all the annealing temperatures, which originates from the change in the exchange coupling in the CoFe/Ru/CoFeB SF pinned layer.[12] When the Ru spacer is thicker than 1.5 nm, no change in the sign of $H_{ex}$ is seen through annealing. On the contrary, when $t_{Ru}$ is less than 1.5 nm, the sign of $H_{ex}$ changes from positive to negative with increasing $T_a$, as shown in the inset of Fig. 2 (a). We believe that this change is caused by the ferromagnetic coupling between CoFe and CoFeB due to thermal diffusion of Ru occurring at high $T_a$, as previously reported,[13] and not by the increase in CoFeB magnetization upon annealing that may change the sign of the effective moment of the pinned layers;[14] the sign reversal was observed only for the samples with thin Ru spacers ($t_{Ru} < 1.5$ nm) and not in the second antiferromagnetically-coupled region ($t_{Ru} > 2.2$ nm). The full recovery of the AP plateau at $T_a = 425$°C suggests direct contact of the CoFe and CoFeB layers as opposed to pin hole formation that leads to partial ferromagnetic coupling. As evident in Fig. 2 (b), no significant change in TMR ratio with respect to the $t_{Ru}$ is observed at $T_a$ = 270°C and 375°C, except for dips due to the incomplete AP configuration arising from weak $H_{ex}$. At $T_a$ = 425°C, the TMR ratio gradually increases as $t_{Ru}$ increases from 0 to 1.5 nm and then saturates when $t_{Ru}$ is more than 1.5 nm. The highest TMR ratio of 361% is observed at $t_{Ru}$ of 2.5 nm. The resistance-area product (*RA*) of all the MTJs falls in the range of 1-4 kΩμm$^2$, which is in good agreement with previous reports.[9,11] There was no correlation between *RA* and $t_{Ru}$.

Figure 3 shows TMR ratio as a function of $T_a$ for S-MTJ and SF-MTJ with $t_{Ru}$ = 1.17 and 2.5 nm. At $T_a$ lower than 325°C, the dependence of the TMR ratio on $T_a$ is similar in all three MTJs. However, when $T_a$ exceeds 325°C, a remarkable difference between S-MTJ and SF-MTJ becomes apparent. The TMR ratio of SF-MTJ increases with increasing $T_a$ and reaches its maximum value of 361% at $T_a$ = 425°C for $t_{Ru}$ of 2.5 nm and 334% at $T_a$ = 400°C for 1.17 nm. In contrast, S-MTJ exhibits its maximum TMR ratio of 181% at $T_a$ = 325°C, and the TMR ratio monotonically decreases with increasing $T_a$, although no significant decrease in $H_{ex}$ was observed (see $t_{Ru}$ = 0 in Fig. 2 (a)); the MR curves of S-MTJs had a flat plateau of the AP state in the measured $T_a$ range, indicating that full AP configuration was achieved.

To understand the difference in the dependence of TMR ratios on $T_a$ for the two types of MTJs (S-MTJ and SF-MTJ), their crystalline structures were examined by HRTEM. Figure 4 shows cross-sectional HRTEM images of S-MTJ ($t_{Ru}$ = 0) and SF-MTJ ($t_{Ru}$ = 2.5); note that $t_{Ru}$ = 2.5 nm resulted in the highest TMR among all the samples. Figures 4 (a) and (b) are respectively the HRTEM images of the S-MTJ and the SF-MTJ after annealing at 270°C. Figures 4 (c) and (d) show the HRTEM images after annealing at 375°C. As can be seen in Figs. 4 (a) and (b), there are no notable differences in the crystalline structure of each layer between the two MTJs at $T_a$ = 270°C, except for the MgO barriers. Both the CoFeB free and pinned layers have amorphous structures and the CoFe have a bcc (110) texture in both samples. The MgO barrier of S-MTJ shows irregularity in the crystalline structure with a weakly (001)-oriented NaCl texture due to the rough surface of the CoFeB pinned layer, whereas that of SF-MTJ shows a clear NaCl (001) structure grown on the flat surface of the CoFeB pinned layer. In spite of the different interface morphology between the CoFeB pinned layers and the crystalline quality of the MgO barriers in the two samples, their TMR ratios do not show much difference up to $T_a$ = 300°C.

After annealing at 375°C, as shown in Figs. 4 (c) (S-MTJ) and (d) (SF-MTJ), the CoFeB free layers of both MTJs crystallized, although their structures were different. The free layer of S-MTJ had a polycrystalline structure with an uncertain texture, whereas that of SF-MTJ had a clear bcc (001) texture. The CoFeB pinned layers of both samples also crystallized into different structures. Analysis with XRD showed that the CoFeB pinned layer of SF-MTJ had a clear bcc (001) texture, which is in good agreement with the TEM images. In contrast, the electron diffraction analysis of the pinned layer of S-MTJ (as shown in the inset of Fig.4 (c)) showed a (110) texture without any distinguishable boundary with the CoFe layer underneath. These results show that the MgO barrier acted as a template for crystallization of both CoFeB free layers and the CoFeB pinned layer in SF-MTJ, whereas crystallization of the CoFeB pinned layer in S-MTJ was dominated by the seeding of the CoFe layer underneath that was in direct contact with the CoFeB layer, resulting in the same texture of the CoFe layer. Inserting a Ru spacer between the CoFe and the CoFeB layers prevents seeding from the bottom CoFe layer, causing the amorphous CoFeB free layer to crystallize from the MgO side. It is also important to note that the Ru spacer decreased the surface roughness induced by the underlayers, leading to formation of an MgO barrier with a clear NaCl (001) texture.

Finally, we discuss the relationship between the crystalline structure and the TMR ratio. When MTJ does not have a Ru spacer, the TMR ratio decreases above 325°C. This can be understood as being due to the crystallization of the CoFeB pinned layer in a bcc (110) texture, which prevents the selective tunneling of the $\Delta_1$ band in the bcc (100) structure that forms the basis of the high TMR ratio.[15-16] This is supported by our earlier study, where we showed that MTJ with a free layer of fcc (111) textured $Co_{90}Fe_{10}(3)$ combined with the pinned layer of CoFe(2.5)/Ru(0.8)/CoFeB(3) (the structure that leads to a bcc (001) pinned layer upon annealing) resulted in a maximum TMR ratio of 131% at $T_a = 350°C$, and its TMR ratio decreased at higher $T_a$, which is similar to the characteristics of S-MTJ in this study.[17] Thus, a possible scenario of having a higher TMR ratio at high $T_a$ in the SF-MTJ with a thicker Ru spacer is as follows. Crystallization of initially amorphous CoFeB electrodes to a (001) bcc structure through annealing causes alignment of the [100] axis of the electrodes to the [110] axis of the crystalline MgO barrier in order to minimize the lattice mismatch[7-8,15-16]; the combination of the (001) MgO barrier with the bcc (001) structure of CoFeB electrodes exhibits a giant TMR ratio as reported in previous studies.[11,16] To withstand a high $T_a$, a corresponding Ru thickness is necessary to prevent the CoFe layer from forming unwanted textures. A thick Ru spacer may also act as a diffusion barrier of Mn from the AF layer to the tunnel barrier.[18-20]

In conclusion, we investigated the relationship between the TMR ratio and the crystallization of the CoFeB layer through annealing in MTJs with highly (001)-oriented MgO barriers, CoFeB free layers, and CoFe/Ru/CoFeB synthetic ferrimagnet (SF) pinned layers with varying Ru spacer thickness. The TMR ratio increased with increasing annealing temperature at larger Ru spacer thicknesses, reaching 361% at the annealing temperature of 425°C. In contrast, the MTJs with pinned layers without Ru spacers exhibited a maximum TMR value of 181% at the annealing temperature of 325°C. Examination using HRTEM showed that the CoFeB pinned layer on the Ru spacer crystallized to a highly oriented bcc (001) structure, and the CoFeB on CoFe crystallized to a bcc (110) structure by annealing. The insertion of the Ru spacer between the CoFe and the CoFeB layers in the SF pinned layer had the effect of forming an (001)-oriented bcc CoFeB pinned layer, resulting in a high TMR ratio through annealing at high $T_a$.

This work was supported by the IT-program of Research Revolution 2002 (RR2002):

Figure captions

FIG. 1. Schematic illustration of sample composed of Si/SiO$_2$/Ta(5)/Ru(50)/Ta(5)/ NiFe(5)/IrMn(8)/CoFe(2.5)/Ru(0.67–2.8)/CoFeB/MgO(1.7)/CoFeB(3)/Ta(5)/Ru(15). The Ru spacer in the pinned layer was deposited to have wedge structure with a varying thickness of 0.67 to 2.8 nm. The S-MTJ sample has no Ru spacer.

FIG. 2. $H_{ex}$ (a) and TMR ratio (b) as functions of $t_{Ru}$ for $T_a$ of 270°C(●), 375°C (▲), and 425°C (□). Data for S-MTJ sample are plotted as $t_{Ru}$ = 0. Antiferromagnetic (AF) and ferromagnetic (F) coupling regions before annealing for each $t_{Ru}$ range are indicated in (a). Inset in (a) shows typical MR curves of the sample for $t_{Ru}$ = 0.83 nm when $T_a$ was 270°C, 375°C and 425°C. Inset in (b) is an MR curve of the SF-MTJ sample when $t_{Ru}$ is 2.5 nm at $T_a$ of 425°C.

FIG. 3. TMR ratios as a function of $T_a$ for S-MTJ (□) and SF-MTJ with 1.17- (■) and 2.5-nm Ru thickness (●). For the SF spin-valve with 2.5-nm Ru, TMR ratio increased up to the very high $T_a$ of 425°C, while that of S-MTJ decreased after 325°C.

FIG. 4. Cross-sectional TEM images of MTJ with CoFe(2.5)/CoFeB(3) (S-MTJ) pinned layer and CoFe(2.5)/Ru(2.5)/CoFeB(3) pinned layer (SF-MTJ); S-MTJ after annealing at 270°C (a) and 375°C (c), SF-MTJ after annealing at 270°C (b) and 375°C (d). All CoFeB layers are amorphous at 270°C, and crystallize differently at 375°C. The MgO barriers have highly oriented NaCl (001) structures. Inset in (c) is the diffraction pattern of the CoFeB pinned layer for S-MTJ after annealing at 375°C.

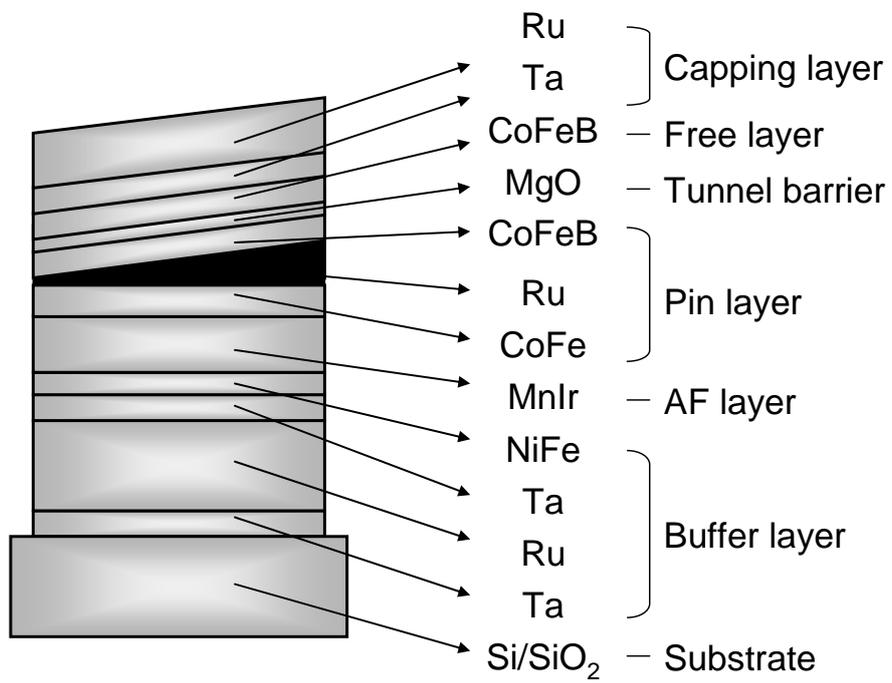

FIG. 1.

Young Min Lee et al.

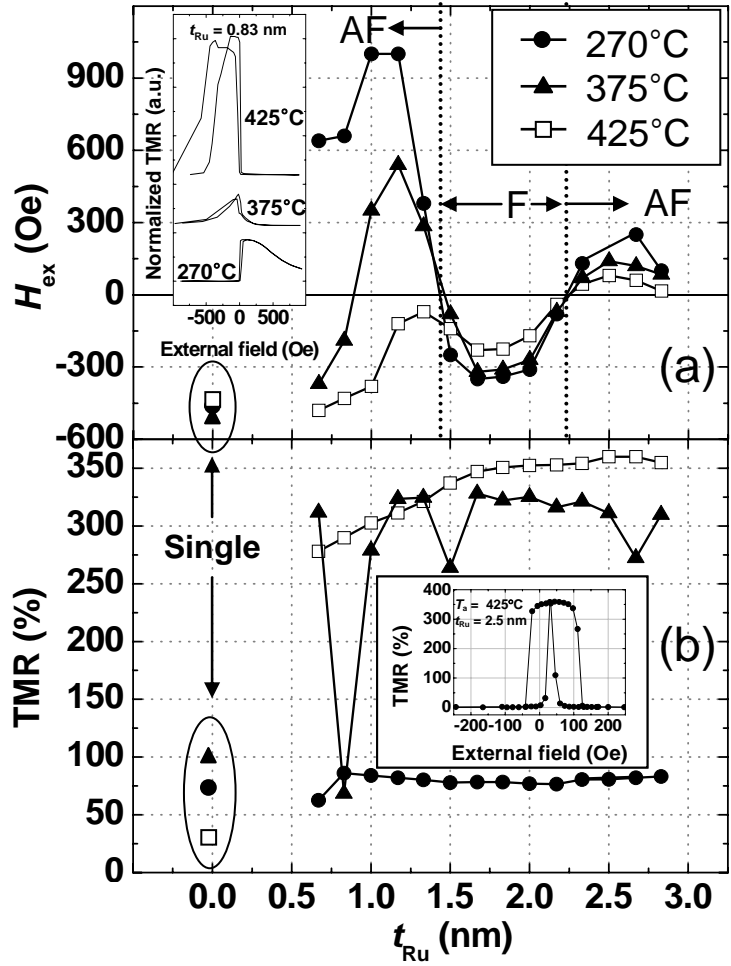

FIG. 2.

Young Min Lee et al.

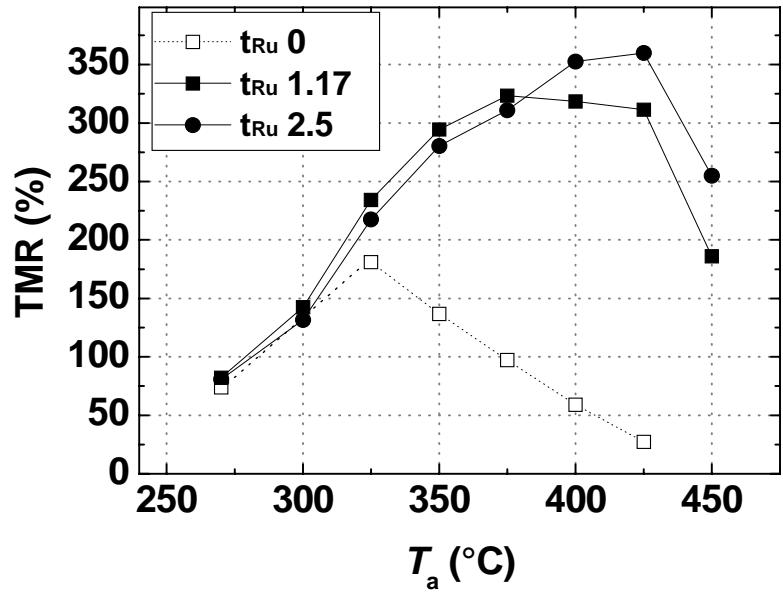

FIG. 3.

Young Min Lee et al.

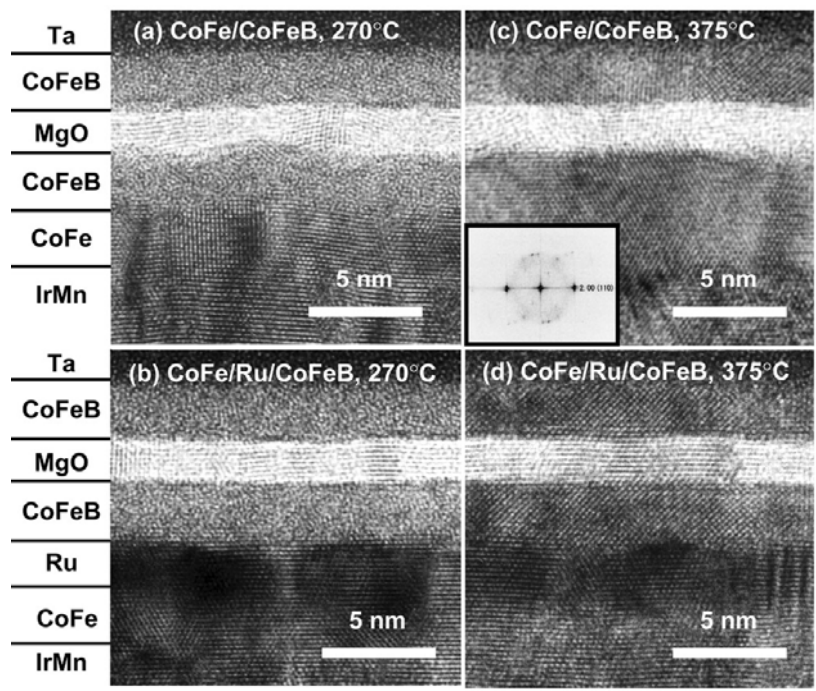

FIG. 4.

Young Min Lee et al.